\documentclass[pra,aps,a4paper,showpacs,twocolumn]{revtex4}
\usepackage{amsfonts}
\usepackage{amsmath}
\usepackage{graphicx}
\usepackage{dcolumn}
\usepackage{bm}

\begin{document}
\title{Schr\"odinger cat states prepared by Bloch oscillation in a spin-dependent optical lattice}

\author{B. J. Wu and J.~M. Zhang}
\affiliation {Beijing National Laboratory for Condensed
Matter Physics, Institute of Physics, Chinese Academy of
Sciences, Beijing 100080, China}

\begin{abstract}
We propose to use Bloch oscillation of ultra-cold atoms in
a spin-dependent optical lattice to prepare schr\"odinger
cat states. Depending on its internal state, an atom feels
different periodic potentials and thus has different energy
band structures for its center-of-mass motion.
Consequently, under the same gravity force, the wave
packets associated with different internal states perform
Bloch oscillation of different amplitudes in space and in
particular they can be macroscopically displaced with
respect to each other. In this way, a cat state can be
prepared.
\end{abstract}

\maketitle

Bloch oscillation is a peculiar response of a particle in a
periodic potential to a weak external force
\cite{bloch,zener}. Under the drag of a constant force $F$,
the wave packet of the particle oscillates back and forth
periodically in space without accelerating indefinitely in
the direction of the force. The weirdness is that a DC bias
generates an alternating current. The underlying reason is
that as long as the force is weak enough, interband
transitions are prohibited by the gaps between energy
bands. Confined in a specific energy band by such a
mechanism, the motion of the particle is captured to a good
extent by the semi-classical equations \cite{kittel}
\begin{eqnarray}
  \frac{d r}{dt}= \frac{\partial E_n(q)}{\hbar \partial q}
  ,\quad   \hbar \frac{d q}{dt}  = F,
\end{eqnarray}
where $r$ and $q$ are the center-of-mass and wave vector of
the particle, respectively, and $E_n(q)$ is the dispersion
relation in the $n$-th band. In the $q$-space, the picture
is that the particle traverses the Brillouin zone, which is
of the topology of a circle, repeatedly at a constant rate.
From these equations, one solves readily the displacement
of the wave packet as
\begin{equation}\label{disp}
    \Delta r (t) = \frac{1}{F} \left[E_n(q_0+ Ft/\hbar) -E_n(q_0)\right],
\end{equation}
where $q_0$ is the initial value of $q$. We see that
$\Delta r$ is a periodic function of time with the period
$T= 2\pi \hbar/a F$---this states the Bloch oscillation in
the semi-classical theory. Here $a$ is the period of the
periodic potential and $2\pi/a$ is the size of the
Brillouin zone. Note that $T$ is independent of the
detailed structure of the periodic potential but depends
only on its period.

Suppose initially the particle is at the bottom of the
lowest energy band ($n=0$) with $q_0=0$. The maximum
displacement $\Delta r_m$ is reached at $t=T/2$, when the
wave vector $q=\pi/a$ arrives at the edge of the first
Brillouin zone. Afterwards, the velocity of the particle
$dr/dt$ reverses. The value of $\Delta r_m$ is simply
\begin{equation}\label{max}
    \Delta r_m = \frac{B}{F},
\end{equation}
where $B\equiv E_0(\pi/a)- E_0(0)$ is the band width of the
lowest band. Simple as it looks, this equation has an
important implication in our work below.

So far, Bloch oscillation has been observed in a variety of
systems, such as semiconductor superlattices
\cite{esaki,Waschke}, cold atoms in optical lattices
\cite{dahan,ferrari,tino,weitz,nagerl,modugno}, and
photonic lattices \cite{pertsch,trompeter,longhi}. On the
application side, it has found use in microwave generation
\cite{esaki,Waschke,grenzer}, precision force measurements
\cite{ferrari,tino,modugno,nagerl}, and coherent transport
of matter waves \cite{alberti,haller,zjm}.

In this paper, we propose that Bloch oscillation can also
be used to prepare Schr\"odinger cat states
\cite{s1,s2,monroe,brune,friedman} in a spin-dependent
optical lattice. The idea is actually very simple. In a
spin-dependent optical lattice, atoms in different internal
states see different potentials (e.g. of different
strengths). This non-trivial fact means that they also have
different energy band structures for their center-of-mass
motion, which in turn means they will have different Bloch
oscillation modes under the same force (see
Eq.~(\ref{disp})). In particular, their maximum
displacements will be different according to (\ref{max})
since the $B$'s may differ, and this implies that the wave
packets corresponding to different internal states will be
displaced with respect to each other. That is, a cat state
can be prepared.

\begin{figure*}[tb]
\begin{minipage}[b]{0.40 \textwidth}
\centering
\includegraphics[ width=\textwidth]{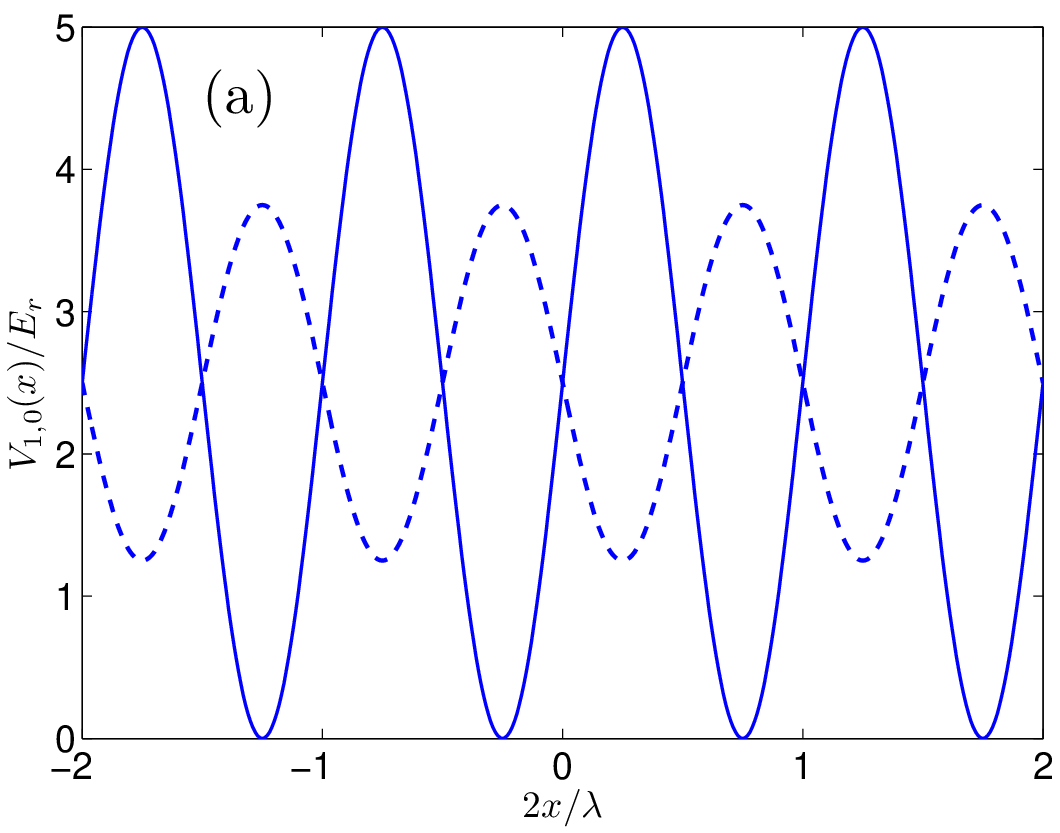}
\end{minipage}
\begin{minipage}[b]{0.40 \textwidth}
\centering
\includegraphics[ width=\textwidth]{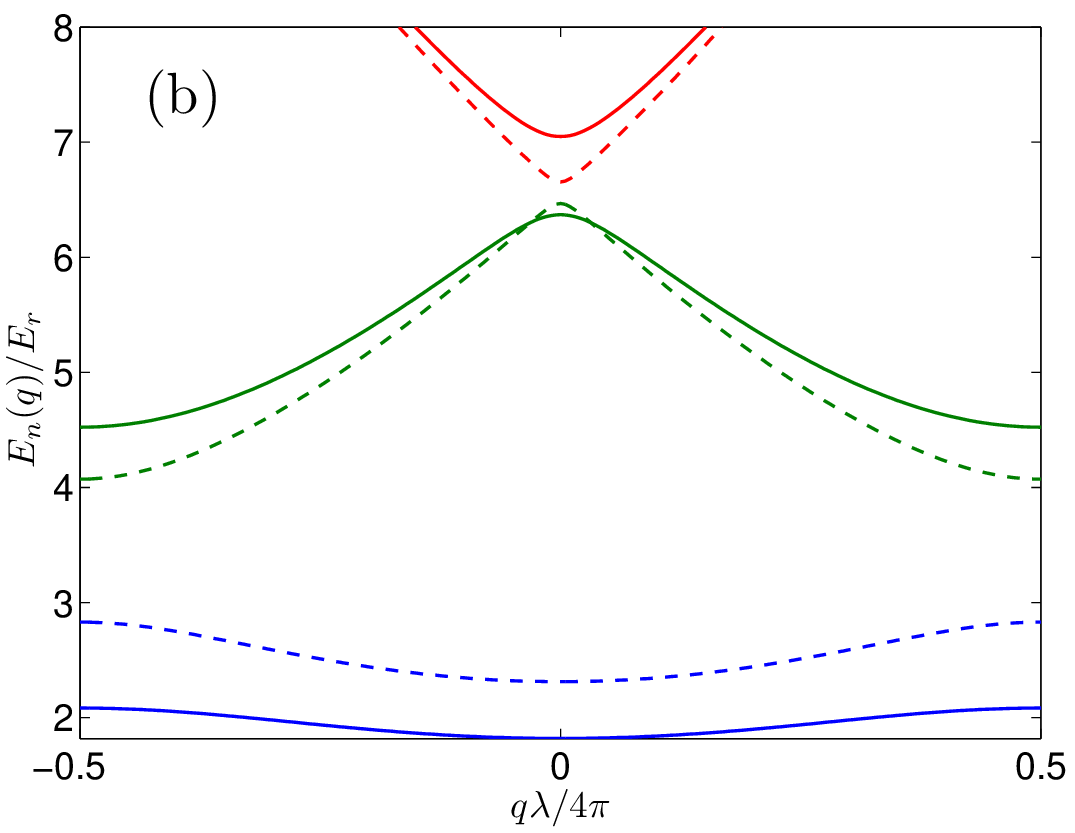}
\end{minipage}
\caption{(Color online) (a) The dipole potentials $V_1(x)$
(solid line) and $V_0(x)$ (dashed line) for an atom in the
internal states $|1\rangle$ and $|0\rangle$, respectively.
The parameters are $V_m/E_r=5$ and $\theta=\pi/2$. Here
$E_r=\hbar^2 k^2/2m$ is the recoil energy of the atom. (b)
The lowest three energy bands for an atom in the
$|1\rangle$ (solid lines) and $|0\rangle$ (dashed lines)
internal states.\label{fig1}}
\end{figure*}

First of all we need a spin-dependent optical lattice. As
proposed in \cite{jessen,jaksch} and realized in
\cite{mandel}, such an optical lattice can be constructed
by interfering two counter-propagating laser beams linearly
polarized but with an angle $\theta$ between the
polarization vectors. The resulting standing light field
can be decomposed into a $\sigma_+$ and a $\sigma_-$
polarized one with intensities $I_+=I_m \cos^2(kx +
\theta/2)$ and $I_-=I_m \cos^2(kx - \theta/2)$,
respectively. Here $k=2\pi/\lambda$ is the wave vector of
the laser beams. Such a decomposition is helpful since the
dipole potential for an atom in a state $|F,m_F\rangle$
(the quantization axis of the atom is along the optical
lattice) is simply the sum of the contributions of the two
components. Below we will use the same system as in
\cite{mandel}. That is, we choose $^{87}$Rb as the atom,
and $|1\rangle\equiv |F=2,m_F=-2 \rangle$ and
$|0\rangle\equiv |F=1,m_F=-1\rangle$ as the two atomic
internal states. The spin dependence of the dipole
potential is realized by choosing the laser frequency to
resolve the fine structure of the Rubidium $D$ line.
Specifically, as in \cite{mandel}, by tuning the wave
length of the optical lattice laser to $\lambda=785$ nm,
the dipole potentials for an atom in the $|1\rangle$ and
$|0\rangle$ states are respectively,
\begin{eqnarray}
  V_1(x;\theta) &=& V_m \cos^2\left(kx-
   \frac{\theta}{2} \right), \quad\quad\quad\quad\quad\quad\quad\quad \nonumber\\
   V_0(x;\theta) &=&  \frac{3}{4}V_m \cos^2 \left(kx+ \frac{\theta}{2} \right) + \frac{1}{4}V_m
   \cos^2 \left(kx-
   \frac{\theta}{2} \right) ,  \nonumber
\end{eqnarray}
where $V_m\propto I_m$. The point is that if $\theta\neq 0$
or $\pi$, $V_1(x)$ and $V_0(x)$ are shifted relative to
each other, and more importantly, have different
amplitudes. The very latter effect results in different
band structures as we see in Fig.~\ref{fig1}. There the
potentials $V_{1,0}$ and the corresponding energy bands for
the two internal states are depicted. The parameters chosen
are $V_m/E_r=5$ and $\theta=\pi/2$, where $E_r=\hbar^2
k^2/2m= 2\pi \hbar \times 3.72$ kHz is the recoil energy of
the atom. Note that the width of the lowest band for the
$|1\rangle$ state is $2\pi \hbar \times 0.983$ kHz, while
that for the $|0\rangle$ state is $2\pi \hbar \times 1.925$
kHz. The two differ almost by a factor of 2.

Now our scheme to generate a Schr\"odinger cat goes like
this. Suppose initially the angle $\theta=0$ (for this
value of $\theta$, $V_1=V_0$) and the atom is in the
$|1\rangle$ state. As for its external state, it is assumed
to be
\begin{equation}\label{ini}
    \Psi_{i} \simeq  \int_{-\pi/a}^{+\pi/a} dq f(q)\phi_0(q).
\end{equation}
Here $\phi_0(q)$ is the Bloch state in the lowest band with
wave vector $q$. The weight function $f(q)$ is localized
around $q=0$ but otherwise unspecified. This condition is
easily satisfied as long as the spatial size of the wave
packet $\Psi_{i}$ is much larger than the lattice constant
$a=\lambda/2$. Actually, the condensate wave function
should satisfy this condition if the condensate is loaded
adiabatically from a magnetic trap into the lattice as is
usually done in cold atom experiments. It is checked that
the results presented in the following are barely affected
with different choices of $f(q)$, as long as the
localization condition is satisfied. This fact is
consistent with the semi-classical theory in which the
details of the wave packets are irrelevant. Specifically,
in the simulations to be presented, $f(q)$ is of the form
$f(q)\propto \exp(-q^2/w^2)$ with $wa/\pi=0.1  \ll 1$. This
value of $w$ corresponds to a wave packet with a size on
the order of $10 a$.

Then at some moment, by using a microwave pulse we can
prepare the internal state of the atom into an arbitrary
superposition of the $|0\rangle$ and $|1\rangle$ states.
Let it be $\alpha |0\rangle + \beta |1 \rangle$ with
$|\alpha|^2 + |\beta|^2=1$. Subsequently, $\theta$ is
adjusted to $\pi/2$ (suddenly or smoothly, it does not
matter; but in our simulation we take the sudden scenario),
and the lattice is tilted by $\varphi$ with respect to the
horizontal plane. The Bloch oscillation then starts. At a
later time $t$, the wave function of the atom is of the
form $\Psi(t) = \alpha |0\rangle \Psi_0(x,t) + \beta
|1\rangle \Psi_1(x,t)$. The evolution of the external wave
function $\Psi_{j} (x,t)$ is given by ($j=0,1$)
\begin{equation}\label{evo}
    i\hbar \frac{\partial}{\partial t} \Psi_{j}  = \left(
-\frac{\hbar^2}{2m}\frac{\partial^2}{\partial x^2} + V_j(x)
-F x \right) \Psi_{j} ,
\end{equation}
with the force $F=mg \sin \varphi$ and the initial
condition $\Psi_j(t=0)=\Psi_{i}$.
\begin{figure*}[tb]
\begin{minipage}[b]{0.30 \textwidth}
\centering
\includegraphics[ width=\textwidth]{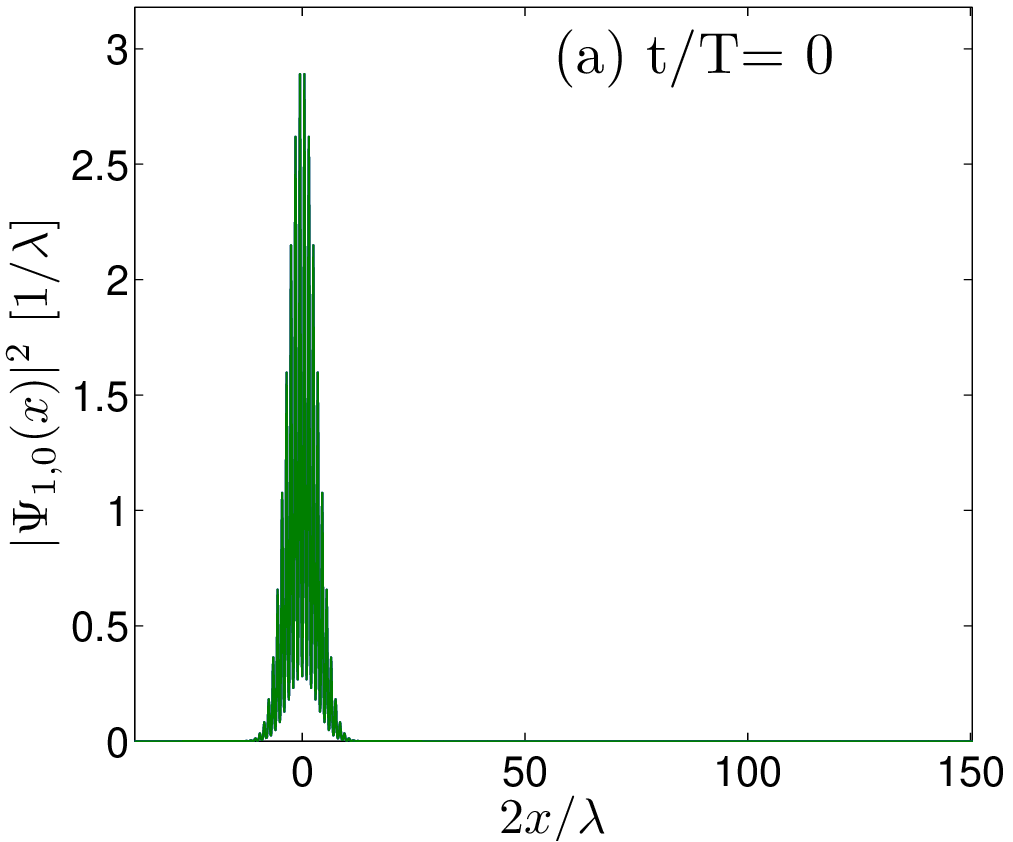}
\end{minipage}
\begin{minipage}[b]{0.30 \textwidth}
\centering
\includegraphics[ width=\textwidth]{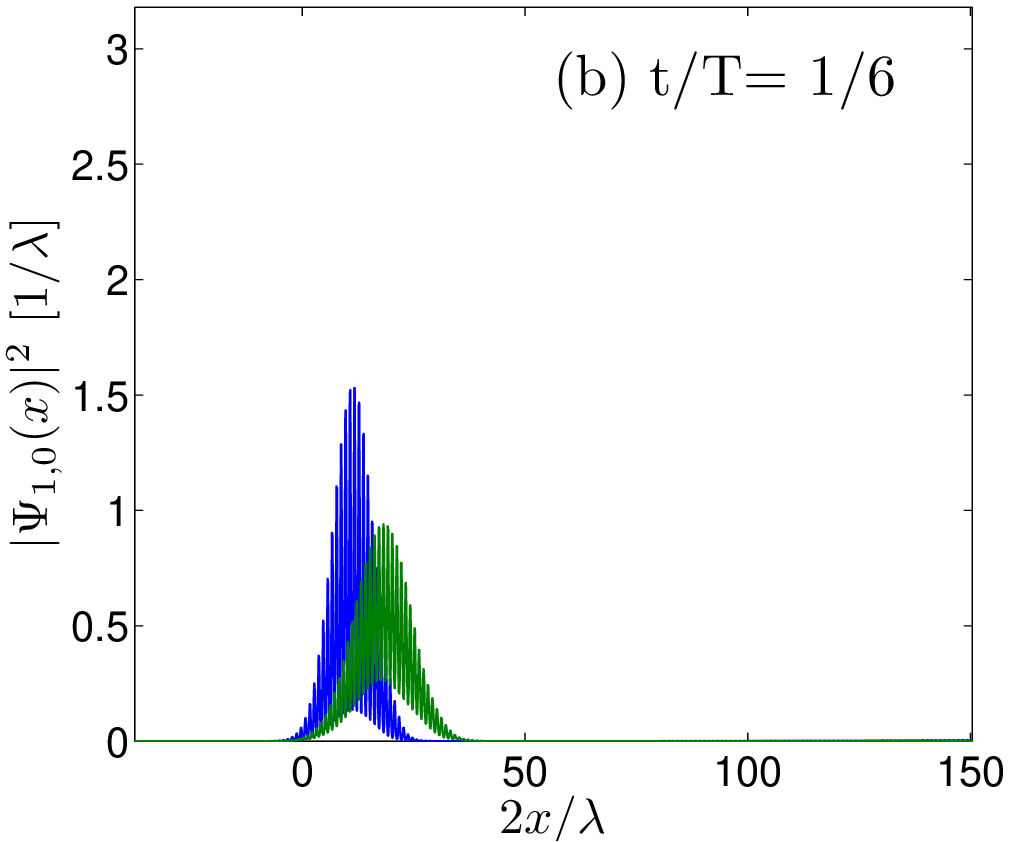}
\end{minipage}
\begin{minipage}[b]{0.30 \textwidth}
\centering
\includegraphics[ width=\textwidth]{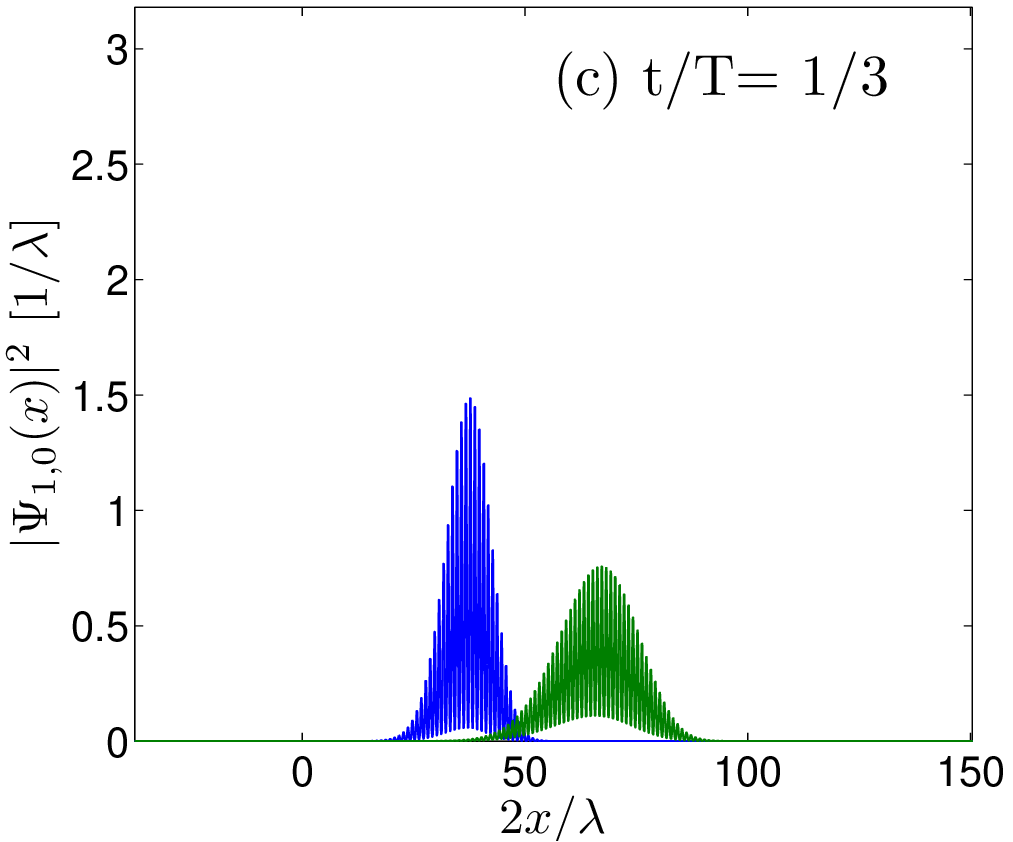}
\end{minipage}

\begin{minipage}[b]{0.30 \textwidth}
\centering
\includegraphics[ width=\textwidth]{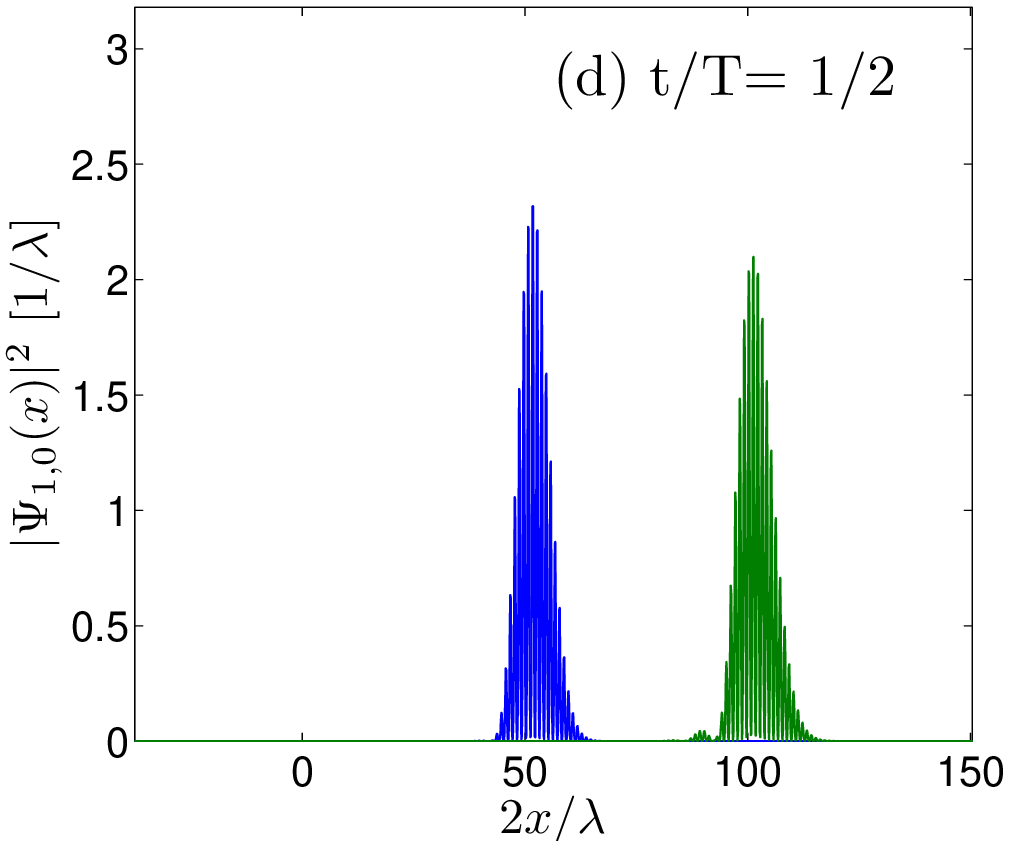}
\end{minipage}
\begin{minipage}[b]{0.30 \textwidth}
\centering
\includegraphics[ width=\textwidth]{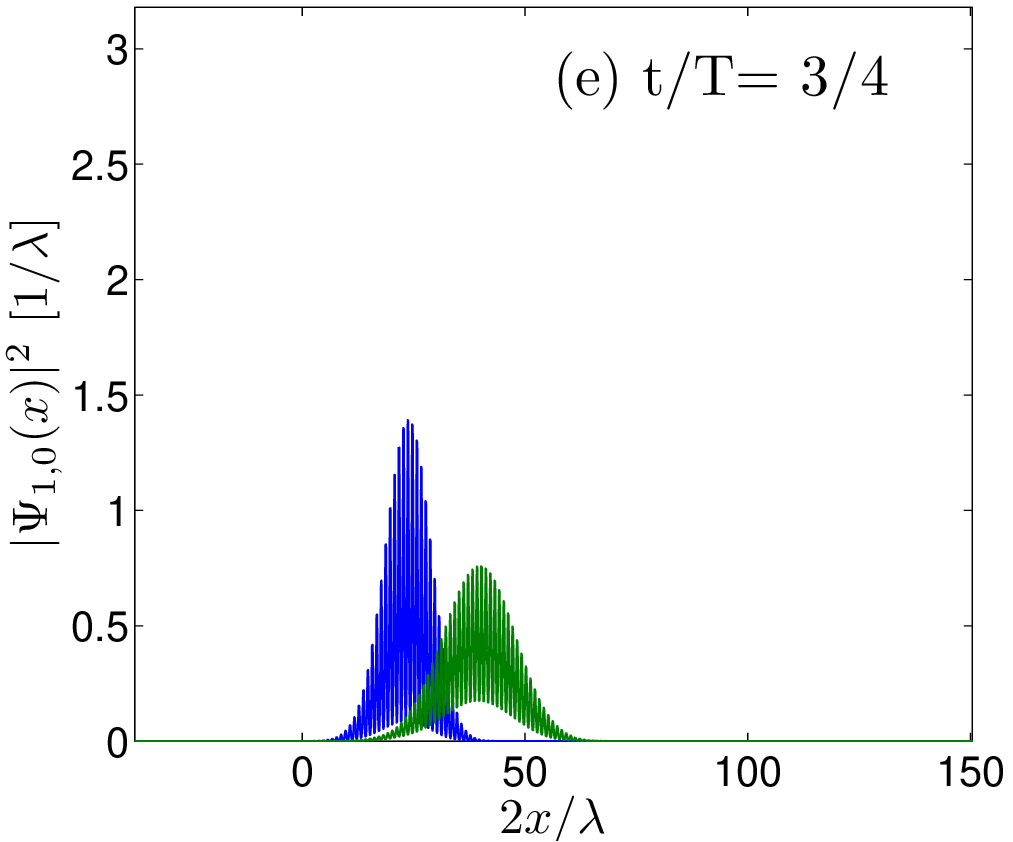}
\end{minipage}
\begin{minipage}[b]{0.30 \textwidth}
\centering
\includegraphics[ width=\textwidth]{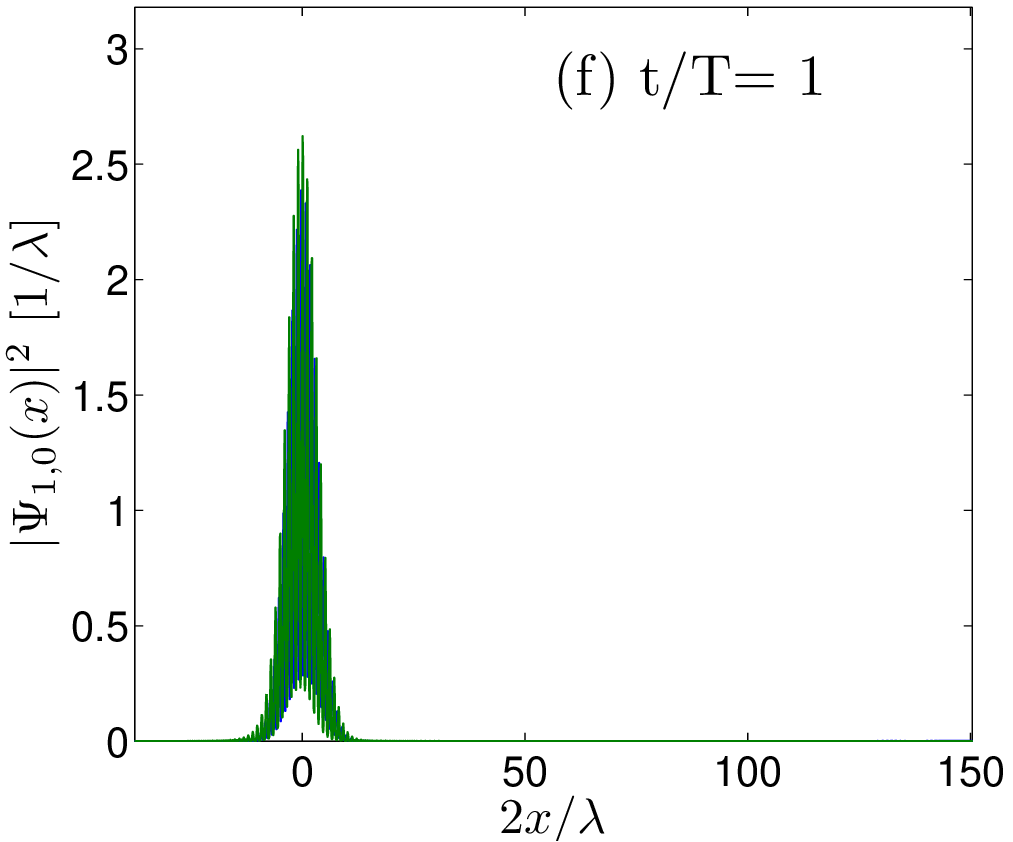}
\end{minipage}
\caption{(Color online) Time evolution of the wave packets
$\Psi_{1}(x,t)$ (blue lines) and $\Psi_{0}(x,t)$ (green
lines) in a Bloch oscillation cycle. The parameters are the
same as in Fig.~\ref{fig1} and $Fa/E_r=0.005$. Initially,
$\Psi_{1,0}=\Psi_i$ as defined in (\ref{ini}). Note that
the wave packets are not normalized to unity.\label{fig2}}
\end{figure*}

Although the semi-classical theory above gives us an
overall idea of the motion of the wave packets
$\Psi_j(x,t)$, here we shall solve Eq.~(\ref{evo})
numerically. Snapshots of $\Psi_j(x,t)$ are shown in
Fig.~\ref{fig2}. As expected, in the interval $0\leq t \leq
T/2$, the two wave packets move rightward and gradually
depart. At the turning point $t=T/2$ (see
Fig.~\ref{fig2}d), the distance between the two reaches the
maximum. Remarkably, at this point, the two wave packets
are well separated---the separation between them is about
$50 a$, which is much larger than their sizes ($\simeq 10 a
$). Note that the wave packets themselves are large enough
to deserve the name macroscopic. Therefore we obtain a
desired Schr\"odinger cat state
\begin{equation}\label{cat}
    \Psi(t=T/2)= \alpha |1\rangle \Psi_1(t=T/2)+ \beta |0\rangle
    \Psi_0(t=T/2),
\end{equation}
for which the internal and external states of the atom are
entangled. The point is that the latter is in two
macroscopically displaced macroscopic states, which
correspond to a ``live'' and a ``dead'' cat, respectively.
The coherence between the two packets can be checked by
applying a microwave pulse to achieve a rotation in the $\{
|1\rangle, |0\rangle \}$ space, and then turning off the
lattice and observing the momentum distribution of the atom
in the $|1\rangle$ state by the absorption imaging method
\cite{mandel}. Note that $\Psi_{1,0}(T/2)$ are both peaked
around $q= \pm \pi/a$ in the momentum space and thus the
interference pattern will primarily consist of two peaks
whose amplitudes depend on the rotation as well as $\alpha$
and $\beta$.

Afterwards, the two wave packets move backwards and return
approximately \cite{recurrence} to their original states at
$t=T$ (see Fig.~\ref{fig2}f). That is, $\Psi_{1,0}(T)
\simeq \Psi_{i}$ up to some global phases \cite{zak}. The
global wave function is then $\Psi(T)\simeq (\alpha
|0\rangle + \beta e^{i\chi} |1\rangle) \Psi_{i}$, where
$\chi$ is the difference of the phases $\Psi_{1,0}$
accumulated in a cycle. Thus the atom completes a Bloch
oscillation cycle by getting its internal state rotated
somehow. In the perspective of the cat state, a cycle of
Bloch oscillation is a cycle of birth-growth-death. The
cycle can be interrupted by putting $F=0$ at an appropriate
time, e.g., at $t=T/2$, when the cat is in its largest
size. Or if the direction of the force $F$ is reversed at
$t=T/2$, the two wave packets will continue moving
rightward instead of going back. This would help to
increase the size of the cat further.

\begin{figure}[tb]
\centering
\includegraphics[width=0.4\textwidth]{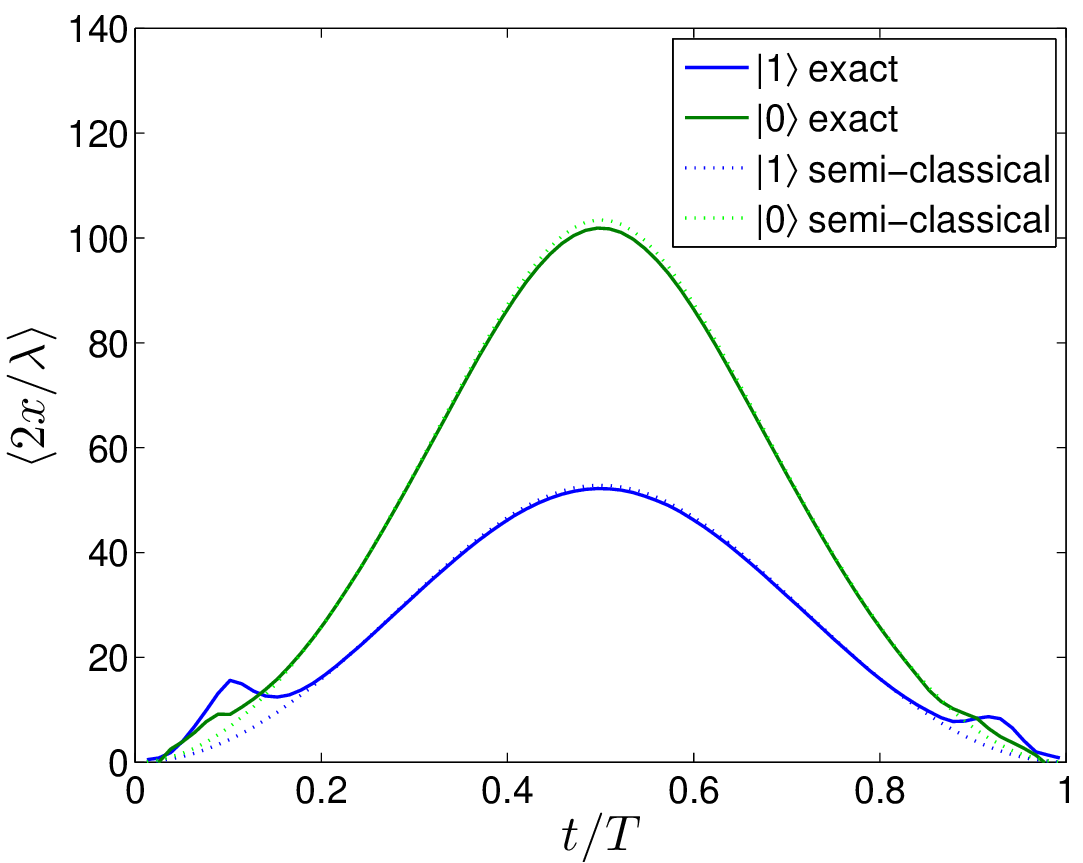}
\caption{\label{fig3} (Color online) Evolution of the
center-of-mass of the wave packets $\Psi_{1,0}(x,t)$ in a
Bloch oscillation cycle. The parameters are the same as in
Fig.~\ref{fig1} and Fig.~\ref{fig2}. The results obtained
by solving Eq.~(\ref{evo}) numerically/exactly (solid
lines) agree very well with those (dotted lines) by the
semi-classical theory [see Eq.~(\ref{disp})]. }
\end{figure}

We have followed the center-of-mass motion of the wave
packets in time. The results are shown in Fig.~\ref{fig3}
(solid lines). We see that the semi-classical theory
(dotted lines) is correct quantitatively. The numerically
exact results deviate significantly from the semi-classical
predictions only in the vicinities of $t=0$ and $t=T$. The
reason is that due to the sudden change of $\theta$, both
wave packets are partially excited and some packets (not
visible in the snapshots) belonging to higher bands are
emitted around $t=0$, which may perform Bloch oscillation
also and return around $t=T$.

We now turn to the problem of the feasibility of the scheme
in experiment. In our simulation, $Fa/E_r=0.005$. It is
essential to make sure that this ratio is much smaller than
unity. First, the potential drop $Fa$ between two
neighboring sites should be much smaller than the gap
between the zeroth and first bands (see Fig.~\ref{fig1}b)
so as to suppress Zener tunneling \cite{zener}. Or
equivalently, the Brillouin zone, especially its boundary,
should be traversed slowly so that transition into excited
bands can be neglected. Second, according to (\ref{max}),
the distance between the two wave packets is inversely
proportional to $F$, thus smaller $F$ means larger
separation or larger ``cat''. Of course, there should be an
optimal value of $F$ since the period $T$ is also inversely
proportional to $F$. For $^{87}$Rb and a lattice constant
$a=\lambda/2=392.5$ nm, the ratio above corresponds to a
tilt angle $\varphi=4^\circ$ and a period $T= 53$ ms. On
the contrary, under the chosen detuning and strength of the
optical lattice, the spontaneous radiation rate
$\Gamma_{eff}$ of the atom is about $0.2$ s$^{-1}$. Thus
the atom is long lived enough to oscillate several cycles
before incoherent processes set in.

In conclusion, we have proposed that the spin-dependent
optical lattice may offer an opportunity to create
Schrodinger cat states by using Bloch oscillation. Our
scheme has several interesting advantages. First, the cat
state experiences birth-growth-death cycles repeatedly. It
would be worthy to study experimentally how this process is
damped in a real optical lattice, which is believed to be
well isolated from the environment. Second, if we start
from a Bose-Einstein condensate and minimize the atom-atom
interaction which is deleterious to the Bloch oscillation,
it might be possible to create a collection of atoms
condensed in a cat state. We note that some generalizations
are also possible. For example, though here we focused on
the one dimensional case, the scheme can be directly
extended to higher dimensions \cite{zjm} since
two-dimensional spin-dependent optical lattices have
already been demonstrated experimentally \cite{sengstock}.
Furthermore, in contrast to the static force considered
here, periodically modulations \cite{alberti,haller} are
worth consideration also since they may help to increase
the size of the ``cat''.

We are grateful to L.~M. Duan, Y. Pan, R.~Q. Wang, and
D.~L. Zhou for stimulating discussions and valuable
suggestions. J.~M.~Z. is supported by NSFC under Grant
No.~11091240226.


\begin{thebibliography}{99}
\bibitem{recurrence}
The recurrence of $\Psi_{i}$ is imperfect since on the one
hand both $\Psi_{1,0}$ are partially excited into higher
bands at the beginning due to the sudden change of
$\theta$, and on the other hand the subsequent evolution is
not strictly adiabatic and higher band components still mix
in \cite{zener}. This feature is not captured by the
semi-classical theory.
\end{thebibliography}
\end{document}